\documentstyle[11pt,fleqn]{article}
\begin{document}

\begin{titlepage}

\begin{flushright}
Freiburg--THEP 94/08\\
April 1994
\end{flushright}
\vspace{1.5cm}

\begin{center}
\large\bf
{\LARGE\bf Two--loop heavy Higgs corrections to the
           Higgs fermionic width}\\[1cm]
\rm
{Adrian Ghinculov}\\[.5cm]

{\em Albert--Ludwigs--Universit\"{a}t Freiburg,
           Fakult\"{a}t f\"{u}r Physik}\\
      {\em Hermann--Herder Str.3, D-79104 Freiburg, Germany}\\[1.5cm]

\end{center}
\normalsize

\begin{abstract}
We calculate the two--loop electroweak corrections to the fermionic
decay width of the Higgs boson at leading order in $m_{H}$.
The sum of one--loop and two--loop
radiative corrections turns out to be at 6\% level
over the whole range of $m_{H}$ where the perturbation theory
is supposed to be a sensible approximation.
We show that the perturbative approach breaks down
at $m_{H} \sim$ 1.4 TeV,
and address the question of its relevance beyond this limit.
\end{abstract}

\vspace{3cm}

\end{titlepage}


That the selfinteraction of the electroweak symmetry breaking sector
of the standard model increases with the mass of the Higgs boson,
and that this eventually leads to the breakdown of the perturbation theory,
belongs to the common wisdom. To establish the limits of perturbation theory
is, however, a more delicate matter.

The coupling constant of the Higgs sector,
$\frac{g}{4 \pi} \frac{m_{H}}{m_{W}}$, becomes of order unity for
$m_{H} \sim$ 1.5 TeV. This is roughly where one expects the
problems related to the asymptotic nature of the perturbative series
to show up.

The perturbative solution for the $S$ matrix is formally unitary,
but its approximations of any finite order are not.
The unitarity violation effects are of higher order in the coupling constant.
If the perturbation theory
fails to converge satisfactorily, they become numerically large,
impairing the relevance of the perturbative amplitudes.
The well--known partial wave analyses of longitudinal vector boson
scattering at tree level by Dicus and Mathur \cite{dicus-mathur} , and
B.W. Lee, C. Quigg and H. Thacker \cite{lee-quigg}
show that the the unitarity bounds are exceeded for $m_{H} \sim$1 TeV.
Durand, Johnson and Lopez \cite{durand1,durand2} extended this analysis
to include the one--loop radiative corrections, and extracted
a bound of $\sim$ 400 GeV. This much smaller value is, however,
to be traced back to their criterion for unitarity violation,
which is stronger than that of refs. \cite{dicus-mathur,lee-quigg},
rather than to genuine one--loop effects.

It is, of course, hard to draw out unambiguous conclusions about
the breakdown of the perturbative series if only its first two
terms are known. The first nontrivial criterion is to compare the
magnitude of the one--loop and two--loop radiative corrections,
and to establish where the series starts to diverge.

Here one must distinguish between radiative corrections to low energy
parameters, and to processes involving the symmetry breaking scalars.
The latter are expected to be larger, since the contributions
which are of leading order in the Higgs mass cancel in the radiative
corrections to low energy parameters
\cite{einhorn:screening,veltman:screening}.
For instance, the two--loop radiative corrections to the $\rho$ parameter
and to the vector boson selfcouplings are quadratic in $m_{H}$,
up to logarithms of $m_{H}$. They become comparable to the one--loop
logarithmic corrections only for Higgs bosons as heavy as 3---4 TeV
\cite{vdBij:2loop:rho,vdBij:2loop:vertex}.

On the other hand, performing multiloop radiative corrections
to processes which involve the Higgs boson is a difficult task.
Already in the case of the two--loop massive diagrams no
general analytical solution exists for nonvanishing external
momenta. One is therefore forced to calculate such diagrams
at least partly numerically.

We calculate in this paper the leading two--loop
corrections to the fermio-nic width of the Higgs boson,
which grow quartically with $m_{H}$. We use the method
described in ref. \cite{2-loop:method} to calculate the
massive two--loop diagrams. The two--loop correction
becomes larger than the one--loop one if $m_{H} >$ 1.4 TeV,
signaling the breakdown of the perturbation theory.
This is compatible with the results of ref. \cite{2-loop:method},
where the two--loop corrections to the shape of the Higgs resonance were
calculated and shown to become large for $m_{H} \sim$ 1.2 TeV.
This is also roughly the value one would expect by considering the
magnitude of the coupling constant of the Higgs sector.
On the other hand, this disagrees with the conclusions of ref.
\cite{maher} about the breakdown of perturbation theory
at 380 GeV \footnote{The two--loop radiative corrections of
ref. \cite{maher} are one order of magnitude larger, which
is also roughly the order of magnitude of the cancellations between
different diagrams. Ref. \cite{maher} probably misses
these cancellations because of a few diagrams evaluated
incorrectly, two of which were already identified in
ref. \cite{2-loop:method}. One also remarks that their $\partial S_{0}$,
which they identify with zero, is a quantity which diverges
logarithmically as the mass of the Goldstone bosons tends to zero.
The proof that $\partial S_{0} = 0$ given in ref. \cite{maher}
is based on illegally splitting this integral into two contributions
whose definition domains do not overlap.}.
In view of this disagreement, special emphasis is put on
checks of both the analytical and numerical parts of the
calculation.

Since we are interested in the leading $m_{H}$ effects,
the natural choice is to work in Landau gauge. Power
counting arguments show that in this gauge only
the Feynman diagrams with scalar propagators survive
\cite{einhorn:screening,marciano}. The diagrams with fermions,
gauge bosons, and Fadeev--Popov fields need not be taken into
account since they are not of leading order. However,
we effectively evaluate the diagrams in a nearly--Landau gauge, that is,
we keep a small, finite gauge parameter $\xi$ throughout the calculation,
and take the limit $\xi \rightarrow 0$ in the final result. This is
necessary at two--loop level in order to avoid the arbitrariness of
$\int d^{n}p \, \frac{1}{p^{4}}$ within the dimensional regularization.
One can convince oneself that this is a regular limit, in the
sense that no finite contributions from the diagrams with gauge or
Fadeev--Popov ghosts survive. This procedure provides also
a good check of the calculation, since all $\xi$ dependence must
cancel in the final results.

Within this framework, the leading corrections to the Yukawa
couplings $- \frac{1}{2} \, i \, g \, \frac{m_{f}}{m_{W}} \, H f \bar{f}$
are a pure renormalization effect. They are simply given by a
factor
$(\frac{Z_{H}}{1-\frac{\delta m_{W}^{2}}{m_{W}^{2}}})^{\frac{1}{2}}$,
so one only needs to evaluate selfenergy diagrams.

Denoting the bare Higgs and Goldstone fields by
$H_{0}$, $w^{\pm}$ and $z^{0}$, one writes down the Higgs sector
of the standard model as:

\begin{eqnarray}
{\cal L} & = &
\frac{1}{2} (\partial_{\mu}H_{0})(\partial^{\mu}H_{0}) +
\frac{1}{2} (\partial_{\mu}z_{0})(\partial^{\mu}z_{0}) +
            (\partial_{\mu}w_{0}^{+})(\partial^{\mu}w_{0}^{-})
                                                \nonumber \\
& & - g^{2}\frac{m_{H_{0}}^{2}}{m_{W_{0}}^{2}} \frac{1}{8} \,
[ \, w_{0}^{+} w_{0}^{-} + \frac{1}{2} z_{0}^{2} + \frac{1}{2} H_{0}^{2}
+ \frac{2 m_{W_{0}}}{g} H_{0}
+ \frac{4 \, \delta t}{g^{2} \, \frac{m_{H_{0}}^{2}}{m_{W_{0}}^{2}}}
 \, ]^{2}
      \; \; ,
\end{eqnarray}
where $\delta t$ is the tadpole counterterm needed to ensure that
the vacuum expectation value of the  Higgs field does not
receive quantum corrections. It is related to the scalar selfenergy
at zero momentum. One further splits the Lagrangian of eq. 1 into a
renormalized Lagrangian and counterterms:

\begin{eqnarray}
H_{0} & = & Z_{H}^{1/2} H      \nonumber \\
z_{0} & = & Z_{G}^{1/2} z      \nonumber \\
w_{0} & = & Z_{G}^{1/2} w      \nonumber \\
m_{H_{0}}^{2} & = & m_{H}^{2} - \delta m_{H}^{2}     \nonumber \\
m_{W_{0}}^{2} & = & m_{W}^{2} - \delta m_{W}^{2}
      \; \; ,
\end{eqnarray}
and fixes the counterterms by using physical, on--shell renormalization
conditions with field renormalization. Note
that the gauge coupling constant $g$ does not get renormalized
at leading order in the coupling constant of the Higgs system
$g^{2} \frac{m_{H}^{2}}{m_{W}^{2}}$.
One can express $g$ in terms of the Fermi coupling constant
and the mass of the charged vector boson as
$g^{2} = 4 \sqrt{2} \, m_{W}^{2} \, G_{F}$, with
$G_{F} = 1.16637 \cdot 10^{-5} \; GeV^{-2}$, and $m_{W} = 80.6 \; GeV$.

The necessary one--loop counterterms at ${\cal O}(\epsilon)$
are given in ref. \cite{2-loop:method}.

One can now proceed with the actual two--loop calculation.
The main task is to evaluate the W boson selfenergy at zero
momentum for extracting the W mass counterterm, and the
on--shell derivative of the Higgs selfenergy for calculating
the field renormalization constant of the Higgs boson. This
involves the selfenergy topologies of. fig. 1. An algebraic
computer program generates all relevant Feynman diagrams,
and uses the techniques of ref. \cite{2-loop:method} to
calculate them. In the case of the selfenergy of the W
boson it is possible to obtain an analytic result because
the external momentum vanishes. For the wave function
renormalization of the Higgs boson, some numerical
integrations have to be performed. The results are:

\begin{eqnarray}
\delta m_{W}^{2 \, (2-loop)} & = & - (g^{2} \frac{m_{H}^{2}}{m_{W}^{2}})^{2} \,
                           (\frac{m_{H}^{2}}{4 \pi \mu^{2}})^{\epsilon} \,
                           \frac{m_{W}^{2}}{(16 \pi^{2})^{2}} \, [ \,
 \frac{3}{32} \frac{1}{\epsilon}
 - \frac{1}{128} + \frac{3}{32} \gamma -
     \nonumber \\
  & & - \frac{\pi^2}{192}
 + \frac{3 \sqrt{3} \pi}{64} - \frac{3 \sqrt{3}}{16} Cl(\frac{\pi}{3})
  \, ]
     \nonumber \\
\delta Z_{H}^{(2-loop)} & = & Re \{ \,
                           (g^{2} \frac{m_{H}^{2}}{m_{W}^{2}})^{2} \,
                           (\frac{m_{H}^{2}}{4 \pi \mu^{2}})^{\epsilon} \,
                           \frac{1}{(16 \pi^{2})^{2}}
 \, [ \, \frac{3}{32} \frac{1}{\epsilon} -
     \nonumber \\
 & &          - 0.416(6) - i \, 1.000(7)
 \, ] \, \} \,
      \; \; ,
\end{eqnarray}
where $\gamma = 0.577215664901532860607$ is the Euler constant,
and $Cl$ denotes the Clausen function, with
$Cl(\frac{\pi}{3}) = 1.01494160640965362502$.

A number of checks were performed
to make sure that these results are correct.

The cancellation of logarithms and poles in $\xi$ was checked
analytically for $\delta m_{W}^{2 \, (2-loop)}$, and
numerically for $\delta Z_{H}^{(2-loop)}$.
The complex integration paths for
$\delta Z_{H}^{(2-loop)}$ were varied to make sure
the integrand is indeed analytical along the integration
path over the Feynman parameters (see ref. \cite{2-loop:method}).
The numerical value of
$\delta Z_{H}^{(2-loop)}$ agrees with the less precise
one which can be derived from the momentum dependence of
the Higgs selfenergy calculated in ref. \cite{2-loop:method}.
Note that the latter result implies numerical integrations
over very different functions, and is sustained by the fact
that its imaginary part agrees with the one--loop
corrections to the Higgs width. The imaginary part
of the momentum derivative of the Higgs selfenergy,
which is given in eq. 3,

\begin{equation}
 - 1.000(7) \, (g^{2} \frac{m_{H}^{2}}{m_{W}^{2}})^{2} \,
               \frac{1}{(16 \pi^{2})^{2}}
     \; \; ,
\end{equation}
obtained by numerically integrating the two--loop diagrams,
agrees with the analytical result which can be derived
by using the Cutkosky rule:

\begin{eqnarray}
- \, (g^{2} \frac{m_{H}^{2}}{m_{W}^{2}})^{2} \,
  \frac{1}{(16 \pi^{2})^{2}} \,
  \frac{3 \, \pi}{4} \,
  \left(
  1 + \frac{\pi \, \sqrt{3}}{12} - \frac{5 \, \pi^{2}}{48}
  \right) & = & \nonumber \\
  = \; \;
 - 1.002245142 \, (g^{2} \frac{m_{H}^{2}}{m_{W}^{2}})^{2} \,
               \frac{1}{(16 \pi^{2})^{2}}
     \; \; .
\end{eqnarray}

Further checks were performed by calculating the two--loop selfenergy
of the Goldstone bosons. After performing an expansion in the
external momentum, one can calculate the necessary
Feynman diagrams analytically. The selfenergy of the Goldstone
bosons at zero momentum agrees with the two--loop tadpole counterterm
given in ref. \cite{2-loop:method}, and the wave function renormalization
satisfies the generalized Ward identity at leading order in $m_{H}$,
$\delta m_{W}^{2} = - m_{W}^{2} \delta Z_{G}$.

We are now in the position to calculate the two--loop
radiative corrections to the Higgs fermionic width:

\begin{eqnarray}
 \frac{Z_{H}}{Z_{G}} & = &
 \frac{1+\delta Z_{H}^{(1-loop)}+\delta Z_{H}^{(2-loop)}
        +\dots}{1+\delta Z_{G}^{(1-loop)}+\delta Z_{G}^{(2-loop)}+\dots}
 \nonumber \\
 & = &
 1 + \delta Z_{H}^{(1-loop)} - \delta Z_{G}^{(1-loop)}
   + \delta Z_{G}^{(1-loop)} ( \, \delta Z_{G}^{(1-loop)} -
 \nonumber \\
 &  &
   - \delta Z_{H}^{(1-loop)} \, )
   + \delta Z_{H}^{(2-loop)} - \delta Z_{G}^{(2-loop)} + \dots
      \; \; ,
\end{eqnarray}
where

\begin{eqnarray}
\delta Z_{H}^{(1-loop)} & = &  g^{2} \frac{m_{H}^{2}}{m_{W}^{2}} \,
                           (\frac{m_{H}^{2}}{4 \pi \mu^{2}})^{\epsilon /2} \,
                           \frac{1}{16 \pi^{2}} \, [ \,
{3\over 2} -
    {{\pi \,{\sqrt{3}}\over 4}} + {\cal O}(\epsilon) \, ]
     \nonumber \\
\delta Z_{G}^{(1-loop)} & = &  g^{2} \frac{m_{H}^{2}}{m_{W}^{2}} \,
                           (\frac{m_{H}^{2}}{4 \pi \mu^{2}})^{\epsilon /2} \,
                           \frac{1}{16 \pi^{2}} \, [ \,
-{1\over 8} + {\cal O}(\epsilon) \, ]
      \; \; .
\end{eqnarray}

Numerically, one obtains:

\begin{equation}
  \frac{Z_{H}}{Z_{G}}  =
  1 +
  .264650 \, \frac{g^{2}}{16 \pi^{2}} \frac{m_{H}^{2}}{m_{W}^{2}}
  - .303(8) \, \left(\frac{g^{2}}{16 \pi^{2}} \frac{m_{H}^{2}}{m_{W}^{2}}
\right)^{2}
  + \dots
      \; \; .
\end{equation}

The results are shown in fig. 2.

Including the two--loop results reduces the magnitude of the radiative
corrections, wich remain under 6\% for the whole range of validity of the
perturbative approach, up to $\sim$ 1.4 TeV, where the
two--loop correction exactely compensates the one--loop one.

Finally, some comments on the possible relevance of perturbation
theory beyond this limit are in order.
In an attempt to sum the asymptotic series of eq. 6,
fig. 3 shows the behaviour of the first nontrivial term of its
Shanks transformation. Since the perturbative series is a power
series, this reduces actually to the Pad\'{e} approximant [1/1].
Certainly, the convergence radius of the perturbative expansion
is zero, and a diagonal sequence of Pad\'{e}
approximants might fail to sum it. To sum the perturbative
series would require knowledge of its high order behaviour.
On the much simpler case of the anharmonic oscillator one
can show that finding the right nonlinear sequence transformation
to sum an asymptotic series depends crucially on the knowledge
of an estimate of the series' remainder \cite{vinette}.
If such information is not available, the only thing one can
do is to compare the results of different summation
recipes \cite{weniger}, but, of course,
knowledge of only the one--loop and two--loop corrections
does not leave much room for experimenting with
nonlinear sequence transformations.

\vspace{.5cm}

{\bf Acknowledgement}

The author wishes to thank prof. Jochum van der Bij for useful discussions.


\newpage


{\bf Figure captions }

\vspace{2cm}

{\em Fig.1}    The topologies of the two--loop selfenergy diagrams.

\vspace{.5cm}

{\em Fig.2}    The radiative corrections $\frac{Z_{H}}{Z_{G}}$
               to the fermionic Higgs
               width in the one--loop (solid line) and
	       two--loop (dashed line) approximations as a function
	       of the mass of the Higgs boson.

\vspace{.5cm}

{\em Fig.3}    The behaviour of the Pad\'{e} approximant
               [1/1] (dotted line) in the
	       strong coupling zone, compared to the
	       one--loop (solid line) and
	       two--loop (dashed line) approximations.


\begin{thebibliography}{99}


\bibitem{dicus-mathur} D.A. Dicus and V.S. Mathur,
                      {\em Phys. Rev. {\bf D7} (1973) 3111.}

\bibitem{lee-quigg}   B.W. Lee, C. Quigg and H. Thacker,
                      {\em Phys. Rev. {\bf D16} (1977) 1519.}

\bibitem{durand1}     L. Durand, J.M. Johnson and J.L. Lopez,
                      {\em Phys. Rev. Lett. {\bf 64} (1990) 1215.}

\bibitem{durand2}     L. Durand, J.M. Johnson and J.L. Lopez,
                      {\em Phys. Rev. {\bf D45} (1992) 3112.}

\bibitem{2-loop:method} A. Ghinculov and J.J. van der Bij,
                      {\em Freiburg-THEP 94/05.}

\bibitem{vdBij:2loop:rho} J.J. van der Bij and M. Veltman,
                      {\em Nucl. Phys. {\bf B231} (1984) 205.}

\bibitem{vdBij:2loop:vertex} J.J. van der Bij,
                      {\em Nucl. Phys. {\bf B255} (1985) 648.}

\bibitem{veltman:screening}   M. Veltman,
                      {\em Acta Phys. Pol. {\bf B8} (1977) 475.}

\bibitem{einhorn:screening}   M. Einhorn and J. Wudka,
                      {\em Phys. Rev. {\bf D39} (1989) 2758.}

\bibitem{maher}       P.N. Maher, L. Durand and K. Riesselmann,
                      {\em Phys. Rev. {\bf D48} (1993) 1061} ;
                      {\em Phys. Rev. {\bf D48} (1993) 1084} ;
                      L. Durand, B.A. Kniehl and K. Riesselmann,
		      {\em DESY 93-131.}

\bibitem{marciano}    W.J. Marciano and S.S.D. Willenbrock,
                      {\em Phys. Rev. {\bf D37} (1988) 2509.}

\bibitem{weniger}     E.J. Weniger,
                      {\em Comp. Phys. Rep. {\bf 10} (1989) 189.}

\bibitem{vinette}     E.J. Weniger, J. \v{C}\'{\i}\v{z}ek and F. Vinette,
                      {\em J. Math. Phys. {\bf 34} (1993) 571.}



\end{thebibliography}
\end{document}